\documentclass[
aps,
reprint,
superscriptaddress,
amssymb,
amsmath
]{revtex4-1}

\usepackage{graphicx}
\usepackage{dcolumn}
\usepackage{bm}
\usepackage{bm, color}
\usepackage{bbm}
\usepackage{lineno}
\usepackage{braket}
\usepackage{amsbsy}

\begin{document}

\title{Coexistence of Weyl semimetal and Weyl nodal loop semimetal phases in a collinear antiferromagnet}

\author{Jie Zhan}
\affiliation{School of Materials Science and Engineering, University of Science and Technology of China, Shenyang, China.}
\affiliation{Shenyang National Laboratory for Materials Science, Institute of Metal Research,Chinese Academy of Sciences, Shenyang, China.}

\author{Jiangxu Li}
\affiliation{Shenyang National Laboratory for Materials Science, Institute of Metal Research,Chinese Academy of Sciences, Shenyang, China.}

\author{Wujun Shi}
\affiliation{Center for Transformative Science, ShanghaiTech University, Shanghai 201210, China}
\affiliation{Shanghai High Repetition Rate XFEL and Extreme Light Facility (SHINE), ShanghaiTech University, Shanghai 201210, China}

\author{Xing-Qiu Chen}
\email{xingqiu.chen@imr.ac.cn}
\affiliation{School of Materials Science and Engineering, University of Science and Technology of China, Shenyang, China.}
\affiliation{Shenyang National Laboratory for Materials Science, Institute of Metal Research,Chinese Academy of Sciences, Shenyang, China.}

\author{Yan Sun}
\email{sunyan@imr.ac.cn}
\affiliation{School of Materials Science and Engineering, University of Science and Technology of China, Shenyang, China.}
\affiliation{Shenyang National Laboratory for Materials Science, Institute of Metal Research,Chinese Academy of Sciences, Shenyang, China.}

\begin{abstract}
Antiferromagnets (AFMs) with anomalous quantum responses have
lead to new progress for the understanding of their magnetic and 
electronic structures from symmetry and topology points of view. 
Two typical topological states are the collinear antiferromagnetic 
Weyl semimetal (WSM) and Weyl nodal loop semimetal (WNLSM). 
In comparison with the counterparts in ferromagnets and non-collinear 
AFMs, the WSMs and WNLSMs in collinear AFMs are still waiting 
for experimental verification. In this work, we theoretically predicted 
the coexistence of Weyl points (WPs) and Weyl nodal loops (WNLs) in 
transition metal oxide RuO$_2$. Owing to the small magnetocrystalline 
anisotropy energy, the WPs and WNLs can transform to each other
via tuning the N\'{e}el vector. Moreover, since the WPs are very close to
Fermi level and the WNLs are even crossing Fermi level, 
the topological states in RuO$_2$ can be easily probed 
by photoemission and STM methods. Our result provides a promising
material platform for the study of WSM and WNLSM states in collinear AFMs.  
\end{abstract}

\maketitle

\section{Background and introduction}
Recently, the understanding of magnetic structures in antiferromagnets (AFMs)
was refreshed according to the symmetry constraint quantum transport properties and 
topological band structures
~\cite{shindou2001orbital,Chen_2014,kubler2014non,vzelezny2014relativistic,
feng2015large,Nakatsuji2015,Nayak2016,Wadley2016,yang2017topological,
vzelezny2017spin,shi2018prediction,ghimire2018large,noky2019linear,
zhangyang2019CrI3,xu2020high,vsmejkal2020crystal,feng2020topological,zhou2021crystal}. 
Generally, a material with magnetic order and zero 
net magnetic moment is considered as an AFM, in which the zero net moment makes it 
not sensitive to external magnetic field. With the progress of understanding in quantum 
responses (such as anomalous Hall effect (AHE) and spin-orbital torque (SOT), et al.), 
some antiferromagnetic structures get intensive interest since they can host
some quantum responses that were believed to only exist in ferromagnetic and
ferrimagnetic counterparts with non-zero net 
magnetizations~\cite{shindou2001orbital,Chen_2014,kubler2014non,vzelezny2014relativistic}.

\begin{figure}[htbp]
\begin{center}
\includegraphics[width=0.48\textwidth]{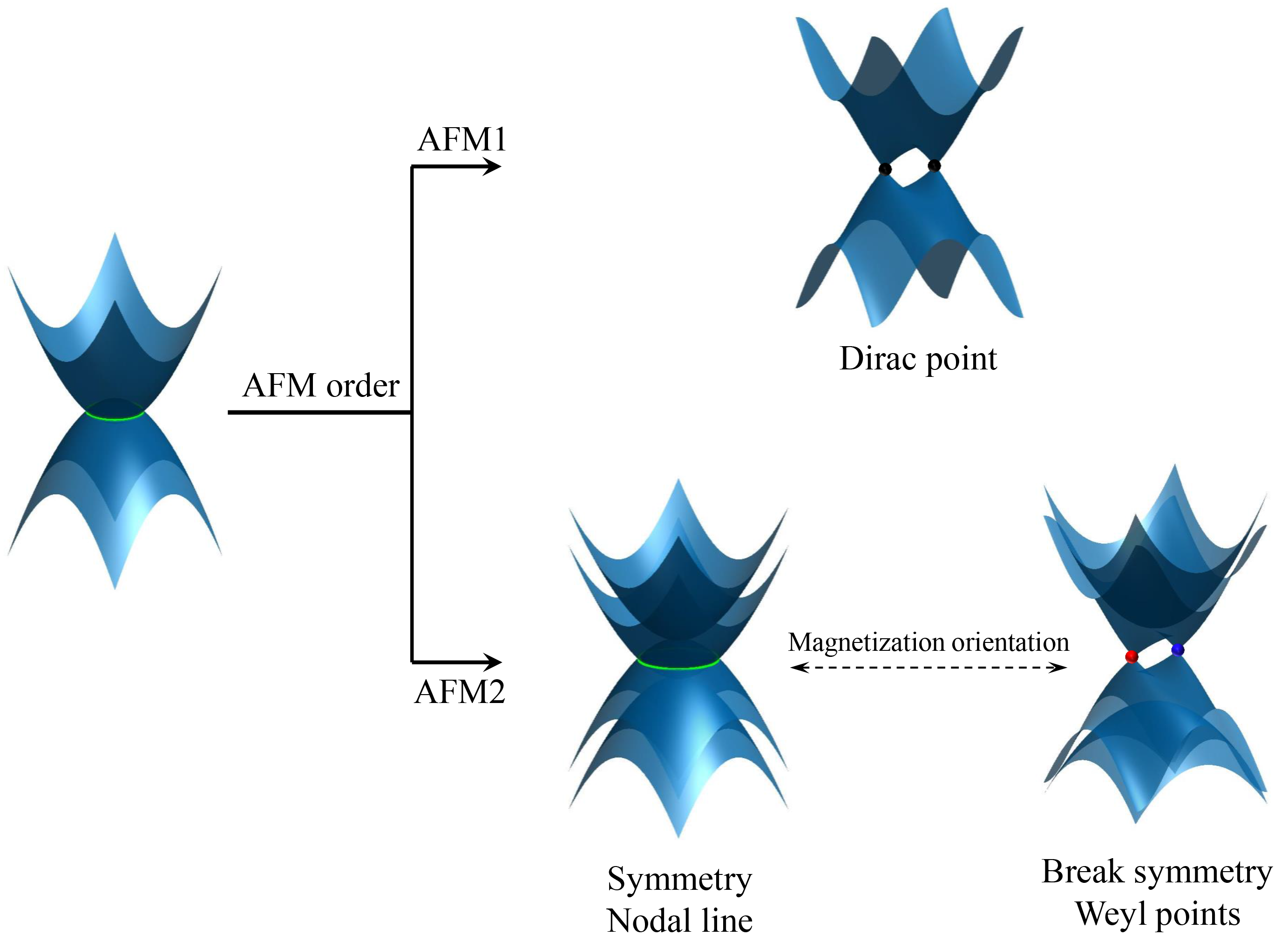}
\end{center}
\caption{
Schematic of possible topological band structures in collinear antiferromagnets (AFMs).
In the case of AFM1 with either $PT$ or $\{T|\tau\}$ symmetry, the linear touching
points of electronic band structure are at least four-fold degenerate in the form of
Dirac points.
In the cases of AFM2 in the absence of $PT$ or $\{T|\tau\}$ symmetries, both doubly degenerate
Weyl points (WPs) and Weyl nodal loops (WNLs) are allowed. Their existence can be controlled via
the magnetization orientation.
}
\label{fig:schematic}
\end{figure}

A typical example is the non-collinear AFM. Though the theoretical prediction of
AHE was as early as 2001~\cite{shindou2001orbital}, its experimental realization
was only achieved after around 15 years later~\cite{Nakatsuji2015}. After that,
the non-collinear AFM gets extensive attention due to their exotic transport properties
and topological band structures.
The observation of AHE~\cite{Nakatsuji2015,Nayak2016},
anomalous Nernst effect (ANE)~\cite{ikhlas2017large,li2017anomalous} and
magneto-optical responses~\cite{higo2018large,liu2018electrical} inspired the
further understanding of collinear AFM. Accordingly, the
collinear AFM can be further classified into two groups that with and
without AHE~\cite{vsmejkal2020crystal}.

Though the existence of intrinsic AHE needs to break                                                                                
time-reversal symmetry, most collinear AFMs can be viewed as two sublattices that
are connected by the joint space group and time-reversal operations. Such 
symmetries can lead to the cancellation of Berry curvature in the whole Brillouin
zone (BZ) and therefore the AHE is forbidden.  Another consequence of the joint 
symmetry is the spin degeneracy in band structures, with which the four-fold 
Dirac points are allowed~\cite{tang2016dirac,vsmejkal2017electric} but that with
lower-fold-degeneracy are forbidden by symmetry, see the up panel of
Fig.~\ref{fig:schematic}

On the other hand, a finite non-zero anomalous Hall conductivity (AHC) is 
symmetrically allowed if such kind of 
joint symmetry is absent and it is usually along with band spin split. It allows
topological band structures with odd-number-fold degeneracy in collinear AFMs, 
where two typical doubly degenerate linear crossing states are Weyl point (WP) and 
Weyl nodal loop (WNL) with non-zero Chern numbers~\cite{wan2011topological,Weng2015,
huang2015weyl,Lv2015TaAs,Xu2015TaAs,fang2015topological}. Together with spin-orbital 
coupling (SOC) and specific magnetization 
orientation, the topological states can be transferred between WPs and WNLs, 
see the bottom panel in Fig.~\ref{fig:schematic}.

With the development of magnetic topological materials,
magnetic WSMs were experimentally verified in ferromagnet
Co$_3$Sn$_2$S$_2$~\cite{liu2018giant,wang2018large,xu2018topological,morali2019fermi,liu2019magnetic}, 
non-collinear AFM Mn$_3$Sn/Ge~\cite{yang2017topological,kuroda2017evidence}, as well as
canted AFM YbMnBi$_2$~\cite{borisenko2019time}.
Besides, magnetic WNLSM was also  directly observed in ferromagnet 
Co$_2$MnGa~\cite{belopolski2019discovery} by ARPES measurements.  

In comparison, despite both WPs and WNLs are allowed in collinear AFMs,
none of them were experimentally observed, so far. The main reason maybe 
due to the complicated metallic electronic structure with several bands
crossing Fermi level and WPs/WNLs far away from Fermi level,
making them hard to detect in experimental measurements~\cite{ghimire2018large}. 
In this work, we find the coexistence of WPs and mirror symmetry protected
WNLs in transition metal oxide RuO$_2$. With WNLs crossing Fermi level and 
WPs only around 0.03 eV bellow Fermi level, it provides a 
model material platform for the direct photoemission and STM  measurements.

\section{Method}
To analyze the topological properties, we calculated the electronic
band structure of RuO$_2$ based on density functional theory by 
full-potential local-orbital code (FPLO) with local basis~\cite{Koepernik1999}. 
Similar to previous calculations, the correlation effect of Ru-4d 
orbitals are considered by the effective on-site Hubbard U, with 
U=2 eV~\cite{vsmejkal2020crystal,zhou2021crystal,dudarev1998electron}.
To calculate the Berry curvature related topological invariants
and surface states, we mapped the Bloch wavefunctions into
symmetry conserving maximally projected Wannier functions~\cite{koepernik2021symmetry} 
and constructed effective tight-binding model Hamiltonians by the 
Wannier functions overlap.

\section{Results and discussion}
\begin{figure}[htbp]        
    \begin{center}              
    \includegraphics[width=0.48\textwidth]{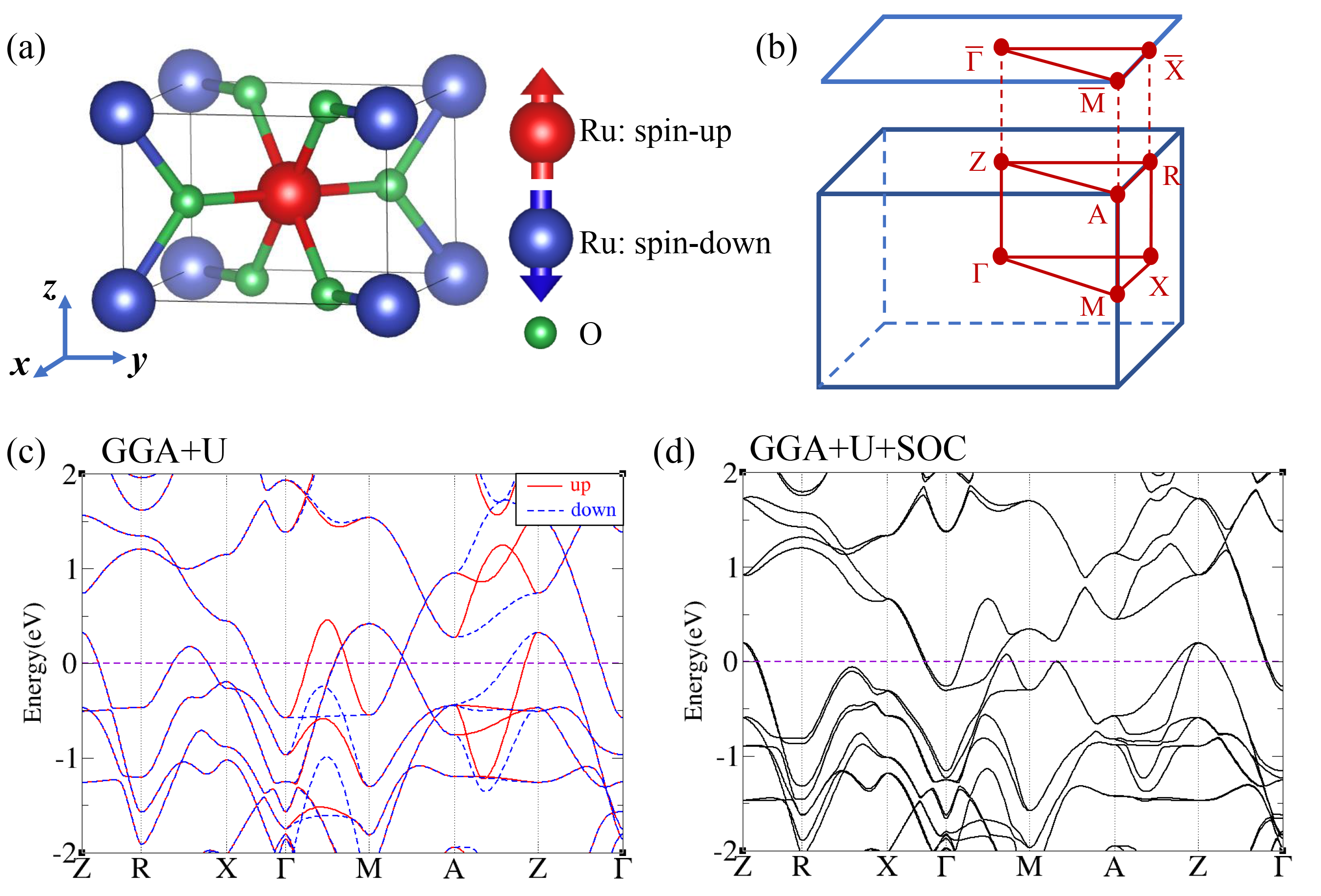}
    \end{center}                
    \caption{                   
    Crystal and electronic structure of RuO$_2$. 
    (a) Magnetic structure of RuO$_2$.
    (b) Three-dimensional Brillouin zone (BZ) for the
    tetragonal crystal lattice and its two-dimensional
    projection in (001) surface.
    (c,d) Energy dispersion of RuO$_2$ without and with 
    the consideration of spin-orbital coupling (SOC).
    }                                                                                                                                        
    \label{fig:lattice}         
\end{figure}

RuO$_2$ is one of the most promising AFM AHE materials          
~\cite{vsmejkal2020crystal,feng2020observation}, in which       
the magneto-optical response can be also manipulated by the orientation
of magnetization and crystal chirality~\cite{zhou2021crystal}.  
It is also the main motivation                                  
for us to choose RuO$_2$ as a typical example to study the WSM and WNL phases
in collinear AFMs. RuO$_2$ has a rutile-type lattice structure belonging to
space group $P4_2/mnm$ (No.136), see Fig.~\ref{fig:lattice}(a). 
The symmetry is reduced by the magnetic order,
depending on the specific spin orientation.   

The magnetization is originated from the $4d$ orbitals on Ru sites,
with easy axis along the $[001]$ direction~\cite{berlijn2017itinerant,zhu2019anomalous}.
Fig.~\ref{fig:lattice}(c-d) shows the energy dispersion
of RuO$_2$ without and with the consideration of SOC. In the 
absence of joint $PT$ symmetry and $\{T|\tau\}$ symmetry 
(with inversion operation $P$, time-reversal operation $T$, 
and fractional translation operation $\tau$) that connect two 
magnetic sublattices, spin-up and spin-down channels are not 
degenerate in the band structure, consistent with previously 
reports~\cite{vsmejkal2020crystal,zhou2021crystal}. Since the two Ru sites are
connected by a screw rotation operation
$S_{4z}=\{C_{4z}|(1/2,1/2,1/2)\}$, the bands from two spin channels
are connected by a $S_{4z}$ operation.

\begin{figure*}[htbp]
    \begin{center}
        \includegraphics[width=0.95\textwidth]{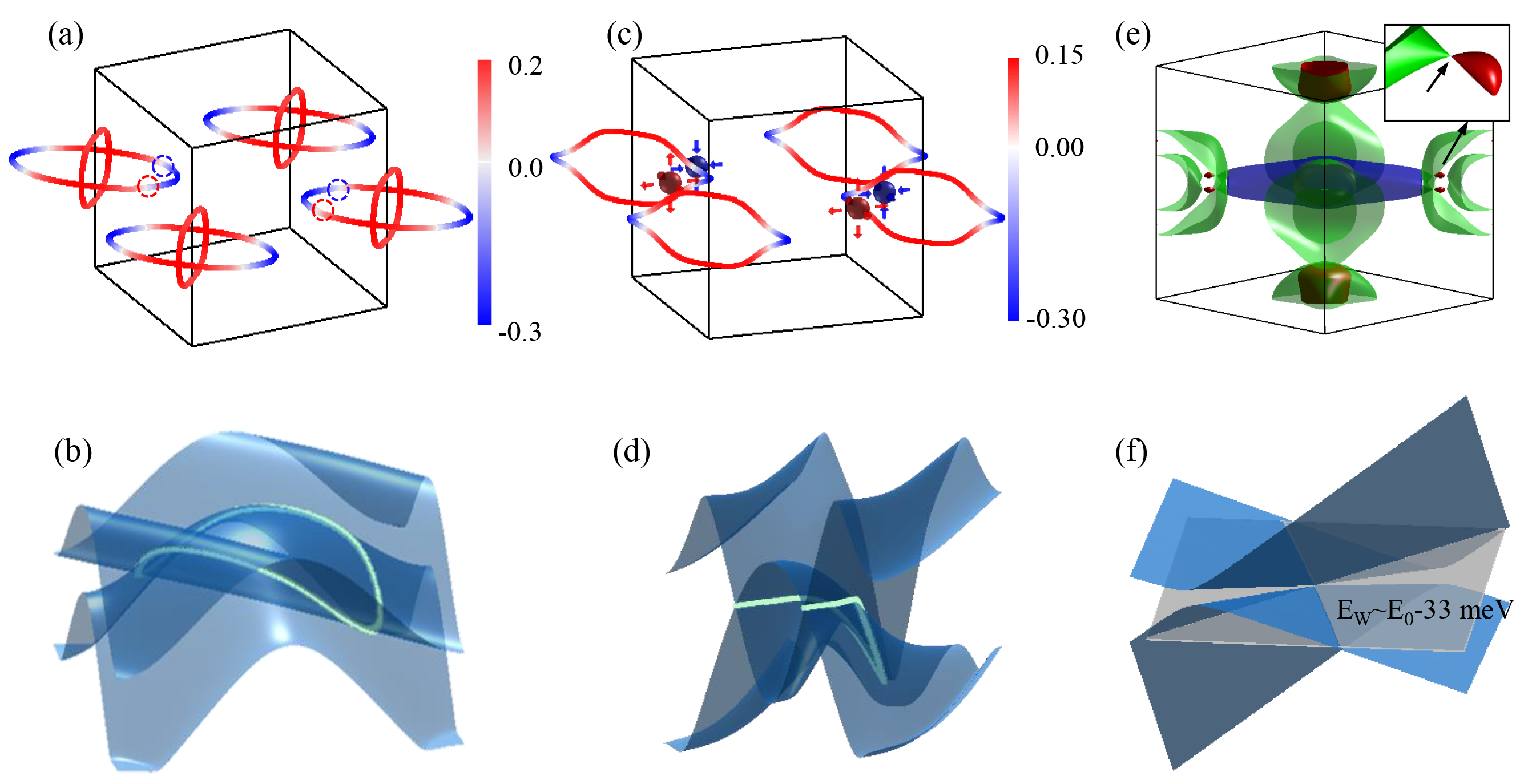}
    \end{center}
    \caption{
      WPs and nodal loops in RuO$_2$.             
      (a) Location of nodal loops from spin-up channel 
          in the case without considering SOC.     
          The dashed circles represent the location of WPs after
          consideration of SOC. The red and blue colors for the 
          dashed circles indicates the chirality of WPs.
      (b) Corresponding three dimensional energy dispersion of the nodal loop in $k_z$=0 plane.
      (c) Location of WPs and WNLs in the case of including SOC,
          with magnetization orientation along [110].        
      (d) Energy dispersion of the WNL             
      (e) Fermi surface of RuO$_2$ with energy lying at $E_W$.
      (f) Energy dispersion around WP.             
      The color bar is in the unit of eV, 0 eV is the Fermi level.          
	}   
    \label{fig:WP}
\end{figure*}

In the case without considering SOC, the system follows the spin rotation
symmetry, and all the mirror symmetries and glide mirror symmetries
are preserved. With the protection of mirror plane $m_{001}$, the band inversion
in both spin-up and spin-down channels can form doubly degenerate nodal
loops in the mirror invariant plane, $k_{z}=0$. As presented in 
Fig.~\ref{fig:WP}(a), the nodal rings in $k_{z}=0$ plane
can extend around half of the BZ in the $k$ space. Moreover, the nodal loops cut 
Fermi level with strong dispersions, where the energy windows are in the 
range of $\sim$-0.3 eV to $\sim$0.2 eV, see the color bar labeled nodal loops
in Fig.~\ref{fig:WP}(a) and three dimensional energy dispersions in Fig.~\ref{fig:WP}(b). 

In addition, the other mirror symmetries of $m_{110}$ and $m_{1-10}$   
also lead to nodal loops in $k_{x}-k_{y}$=0 and $k_{x}+k_{y}$=0 planes, 
respectively, and they touch the nodal loops in $k_{z}$=0 plane on the 
$\Gamma$-M line. But the nodal loops in $k_{x}\pm k_{y}$=0 are around 
0.15 eV above Fermi level and not easy to detect in experiments.      
Owing to the screw rotation operation that connects two Ru-sites,  
the nodal loops from spin-up channel around the $(-\pi,\pi,0)$ corner 
and that from spin-down channel around the $(\pi,\pi,0)$ corner are
linked by a $C_{4z}$ rotation operation.   

With the inclusion of SOC, the existence and behavior of the original nodal 
rings depend on the specific magnetization orientations, see the sketch in 
Fig.~\ref{fig:schematic}. When the N\'{e}el vector is perpendicular to the
plane, the original nodal loops can be preserved if the
wavefunctions of the two bands have opposite mirror 
eigenvalues. Otherwise, the gapless linear crossing will be broken
with opening band gaps, and WPs are allowed nearby. Meanwhile, 
when SOC is taken into consideration, the hybridization of 
two spin channels can generate additional WNLs and WPs.

We noted that previous $ab-initio$ calculations found the magnetocrystalline anisotropy
energy is only around 2.76 meV/Ru~\cite{zhou2021crystal}. Such
small energy difference illustrates the easy tunability of the N\'{e}el vector.
Weak doping in the form of Ru$_{1+x}$O$_{2-x}$ and Ru$_{1-x}$Ir$_x$O$_2$ can
change the N\'{e}el vector to lying within (001) plane~\cite{vsmejkal2020crystal}.
Recently experimentally grown $[001]$ oriented
RuO$_2$ thin films have the in-plane magnetization~\cite{feng2020observation}.
Therefore, most studies about the electrical and optical responses focused on 
the cases with in-plane N\'{e}el vector.  With tiny hole-doping, the AHC
can reach up to $\sim 300$ S/cm~\cite{vsmejkal2020crystal},
a big value in all AMFs. In linear
magneto-optical response, both Kerr rotation angle and Faraday rotation
angle increase along with rotating N\'{e}el vector away from $[001]$ axis
to (001) plane, and the peak value can reach up to $\sim$0.6 and $\sim$2.0 degree,
respectively~\cite{zhou2021crystal}.

Considering the strong non-trivial quantum responses of RuO$_2$ mainly happen in the 
cases with magnetization in (001) plane~\cite{vsmejkal2020crystal,zhou2021crystal},
we will also focus on the same situations.
Taking magnetization along $[110]$ as an example, the magnetic structure breaks 
most of the mirror symmetries, and only $m_{110}$ is left. Therefore the WNLs are
only allowed in the (110) plane, namely, in the planes satisfying $k_{x}+k_{y}=0$ 
or $k_{x}+k_{y}=\pi$.

As presented in Fig.~\ref{fig:WP}(c), one independent WNL exists in the $k_{x}+k_{y}=0$
plane, which can extend around half of the BZ. With doubly degenerate linear crossing,                                                         
the WNL hosts a $\pi$ Berry phase for the Wilson loop around it.
Nearby the hinge of BZ with $k_{x}=k_{y}=\pi$, the WNL is almost flat with very weak 
dispersion. A sharp dispersion appears away from the hinge, with energy window in 
the range of -0.3 eV to 0.15 eV, crossing the Fermi level.  
This WNL is constructed by the bands with opposite spin orientations
and does not exist in the situation without considering SOC.

In addition to the WNLs, we also find one pair of independent WPs with 
opposite chiralities in $k_z=0$ plane. The WPs are originated from the nodal 
rings in Fig.~\ref{fig:WP}(a) in the case without including SOC, see the 
red and blue dashed circles. The $[110]$ oriented magnetization together 
with SOC break the original mirror symmetry $m_z$. Hence, the gapless nodal 
loops were broken with opening band gaps. Meanwhile, the WPs 
are formed on the the original nodal loops, similar to the situation in 
Co$_3$Sn$_2$S$_2$~\cite{liu2018giant,wang2018large,xu2018topological}.
Considering the inversion symmetry, there are four WPs 
near Fermi level in the 1st BZ in total, see Fig.~\ref{fig:WP}(a).   
Since the location of WP is close to the Fermi cutting between the 
original nodal rings and Fermi energy, this pair of WPs are very close 
to Fermi level, with an energy of $E\sim E_{0}-0.03$ eV,
making them very easy to detect by the surface measuring techniques,                                                                        
such as ARPES and STM. 
We further checked the Fermi surface by fixing the energy at WP
and found that it also presents as a linear touching of the Fermi
surfaces. As shown in Fig.~\ref{fig:WP}(e), there are four Fermi surfaces
at $E=E_{W}$, consistent with the energy dispersion plotted in Fig.~\ref{fig:lattice}(d). 
Owing to the $PT$ and $\{T|\tau\}$ symmetry breaking and
strong local magnetization on Ru sites, all the Fermi surfaces are 
with large spin split, except for some high symmetry points and the 
special doubly degeneraced WPs. As presented in the upper-right corner of 
Fig.~\ref{fig:WP}(e), the WP presents as a linear touching of 
one small red bubble and one large green bubble from original 
spin-up channel. 

Correspondingly, the energy dispersion 
is strongly tilted near the WPs. The constant energy plane 
at $E_{W}$ cuts both electron and hole bands through the 
linear crossing point, see Fig.~\ref{fig:WP}(f). Therefore, 
it is a typical type-II WP~\cite{Soluyanov2015WTe2,Sun2015MoTe2}. 
Rotating the N\'{e}el vector from $[110]$ to $[100]$ in 
(001) plane or from $[110]$ to $[001]$ in (1-10) plane, 
the positions of WPs are also shifting in both 
energy and $k$ spaces. In some specific angles, they are 
canceled out in the situation of meeting the other WP 
with opposite chirality.

\begin{figure}[htbp]
    \begin{center}
        \includegraphics[width=0.48\textwidth]{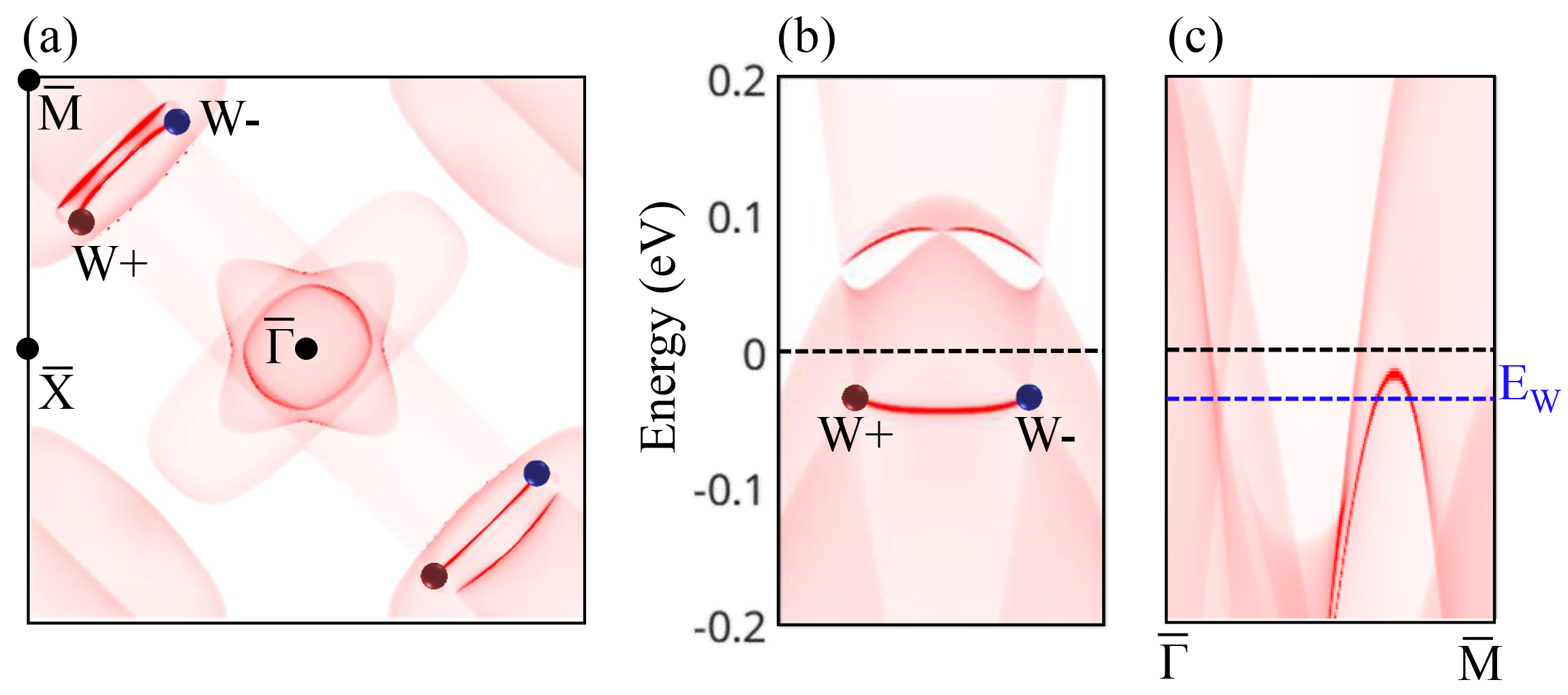}
    \end{center}
    \caption{
    Surface states of RuO$_2$.
    (a) Fermi surface with energy lying at WP.
    (b) Energy dispersion crossing one pair of WPs with opposite
        chirality.
    (c) Energy dispersion along $\overline{\Gamma}-\overline{M}$.
        Red and white color demonstrate the occupied and unoccupied states,
        respectively.
    }
    \label{fig:surface}
\end{figure}

A typical feature of the WSM is the existence of non-closed
topological surface Fermi arc~\cite{wan2011topological}. 
To calculate the surface state, we considered a half-infinite 
open boundary condition along $[001]$ direction by the 
Green's function method~\cite{Sancho1984, Sancho1985}. 
Putting the energy at the $E_W$, one can easily see 
a Fermi arc surface state connecting two WPs with
opposite chiralities, see Fig.~\ref{fig:surface}(a). Though 
the bulk magnetic structure of RuO$_2$ has inversion symmetry, 
the top or bottom surfaces does not follow it after cutting 
bulk chemical bonding. Hence, these two Fermi arcs near 
$(-\pi/a,\pi/a)$ and $(\pi/a,-\pi/a)$ are not connected
by any symmetries and their detailed shapes are different.
The Fermi arc states are further confirmed by the energy 
dispersion along the line crossing one pair of WPs. 
The surface band starts from the bulk linear crossing point 
(W+) and ends at the other one (W-), see Fig. ~\ref{fig:surface}(b).

In addition to the above Fermi arc, one can see the other surface state
almost parallel to the Fermi arc, which merges into bulk states at
the ends. At first glance, it also shows a Fermi arc feature.
To see the physical origin of this surface state, we plot the surface
energy dispersion along $\overline{\Gamma}-\overline{M}$, by crossing
the connection of two WPs. As presented in Fig.~\ref{fig:surface}(c), 
both of them indeed belong to the same surface band and the constant energy
$E_W$ cuts this band twice. Therefore, both of them are parts of the Fermi arcs. 
In principle, there should be an odd number of Fermi crossings with only one 
pair of WPs. But because of the complicated bulk band overlap, there is not an 
indirect band gap between $\overline{\Gamma}-\overline{M}$ and some
surface states are merged into bulk, making the other Fermi crossings
not visible.

\section{summary}
In summary, we systematically understand the topological band structure
in collinear AFM RuO$_2$ from first-principle calculations and symmetry analysis. 
In absent of joint $PT$ and $\{T|\tau\}$ symmetries, WPs and WNLs 
can coexist in RuO$_2$ and they can transform to each other via 
magnetization orientation. Furthermore, the WPs are only around 30 meV below 
Fermi level and the WNLs cut Fermi level. After the discovery of 
WSMs and WNLSMs in ferromagnets and non-collinear AFMs, RuO$_2$ 
provides a promising material platform for direct observation 
of WPs and WNLs in collinear AFMs.

\begin{acknowledgments}
This work was supported by the National Science Fund (Grant No. 51725103, 52188101) 
and Jiangxu Li acknowledges the support from the fellowship
of China Postdoctoral Science Foundation (Grant No. 2021M700152).
\end{acknowledgments}

\bibliography{transport.bib}

\end{document}